\documentclass{iaus}
\usepackage{graphicx}

\title[Environment and upper limit to massive stars] 
{Environments of massive stars\\ and the upper mass limit}

\author[Paul A. Crowther]   
{Paul A. Crowther}

\affiliation{Department of Physics and Astronomy, University of Sheffield, \\
Hounsfield Road, Sheffield, United Kingdom, S3 7RH \\ email: {\tt Paul.Crowther@sheffield.ac.uk} }

\pubyear{2012}
\volume{279}  
\pagerange{9-17}
\jname{Death of Massive Stars: Supernovae and Gamma-Ray Bursts}
\editors{P. Roming, N. Kawai \& E. Pian, eds.}
\begin{document}

\maketitle

\begin{abstract} The locations of massive stars ($\geq 8 M_{\odot}$) 
within their host galaxies is reviewed. These range from distributed 
OB associations to dense star clusters within giant 
H\,{\sc ii} regions. A comparison between massive stars and the environments of 
core-collapse supernovae and long duration Gamma Ray Bursts is made, both at
low and high redshift. We also address the question of the upper stellar 
mass limit, since very massive stars (VMS, $M_{\rm init} \gg 100 
M_{\odot}$) may produce exceptionally bright core-collapse supernovae or 
pair instability supernovae. \keywords{stars: early-type, stars: 
supernovae: general, stars: Wolf-Rayet, ISM: HII regions, galaxies: star 
clusters, galaxies: ISM} \end{abstract}

\firstsection 

\section{Environments of Massive Stars}

Massive star formation in the Milky Way spans a broad spectrum, 
from dispersed, low intensity OB associations to concentrated, high
intensity starbursts. Within a few hundred parsec of the Sun, high mass 
stars ($M_{\rm init} \geq 8 M_{\odot}$) are rather distributed, typically 
located in loose, spatially extended OB associations (de Zeeuw et al. 
1999). A notable  exception is Orion OB1, which hosts the Orion Nebula 
Cluster (ONC),  responsible for our closest H\,{\sc ii} 
region. Further afield, large numbers of 
massive stars are  associated with relatively intense bursts of star formation
such as the high mass, compact clusters (Trumpler 14, 16) within the
Carina Nebula giant  H\,{\sc ii} region.

\subsection{Star clusters}

It is generally accepted that the majority of stars form 
within star clusters (Lada \& Lada 2003), although recent evidence 
suggests star formation occurs in a continuum of stellar densities (e.g. 
Evans et al. 2009).  Nevertheless, given their short-lifetimes (3--50 Myr) 
only a few percent of massive stars appear genuinely `isolated' (de Wit et 
al. 2005) such that they either tend to be associated with their natal cluster or 
are plausible runaways from it\footnote{Runaways may be ejected from their 
cluster either dynamically during the formation process or at a later 
stage after receiving a kick following a supernova explosion in a close 
binary system.}.

According to Weidner \& Kroupa (2006), there is a tight relation 
between cluster mass, and the most massive star formed within the cluster, 
although this remains controversial (Calzetti et al. 2010, Eldridge 2012).
Examples of well known star clusters spanning a range of masses are shown in 
Table~\ref{table1}, all of which are sufficiently young ($<1-2$ Myr) that 
the most massive stars have yet to end their lives. We include the most 
massive star in each cluster, which increases towards the highest mass 
clusters.

If there is a relation between a cluster and its most massive star, the 
galaxy-wide stellar initial mass function (IMF) will also depend upon the 
cluster mass function and the range of cluster masses, as set out by 
Pflamm-Altenburg et al. (2007). In normal star-forming galaxies the 
cluster mass distribution follows a power law with index $-$2, 
albeit this is truncated at high mass depending upon the rate of star formation
(Gieles 2009). Consequently, similar absolute numbers of stars are formed 
in low mass ($M_{\rm cl} \sim 10^{2} M_{\odot}$), intermediate mass ($\sim 
10^{3} M_{\odot}$)  and high mass ($\sim 10^{4} M_{\odot}$)  clusters, 
while high mass stars should be rare in the former. This is not always the case, 
since star formation in some nearby dwarf irregular starbursts is strongly
biased towards a few very high mass clusters (e.g. NGC 1569, Hunter et al. 2000).

\begin{table}
  \begin{center}
  \caption{Selected young star clusters spanning a range of masses, 
 $M_{\rm cl}$, 
for which (initial) masses of the highest mass stars, $M_{\ast, init}$, 
 have been determined.}
  \label{table1}
  \begin{tabular}{crcccc}\hline 
  Cluster & $M_{\rm cl}/M_{\odot}$ & Ref & Star & $M_{\ast, 
 init}/M_{\odot}$ & Ref \\
  \hline
 $\rho$ Oph & $\sim\,10^{2}$ & a & $\rho$ Oph Source 1 & 9 &  a\\
 ONC  & $1.8 \times 10^{3}$ & b & $\theta^{1}$ Ori C & 39$\pm$6 & c \\
NGC 3603 (HD 97950)   & $\sim\,10^{4}$  & d & NGC 3603-B & 166$\pm$20 & e 
 \\
 R136 (HD 38268) & $5 \times 10^{4}$ & e & R136a1   & 320$^{+100}_{-40}$ 
 & e \\
  \hline
  \end{tabular}   
  \end{center}
 \scriptsize{
  (a) Wilking et al. (1989);  (b) Hillenbrand \& Hartmann (1998);  (c) 
Sim\'{o}n-D\'{i}az et al. (2006) \\
  (d) Harayama et al. (2008); (e) Crowther et al. (2010)}
\end{table}

\subsection{H\,{\sc ii} regions and star formation rates}

Because of the (universal?) Salpeter IMF slope, the overall statistics of 
massive stars in galaxies will be heavily biased towards 8--20 $M_{\odot}$ (early 
B-type) stars. However, the most frequently used indicator of active 
star formation is nebular hydrogen emission (e.g. H$\alpha$) from gas 
associated with young, massive stars. The Lyman continuum ionizing output 
from hot, young stars is a very sensitive function of temperature (stellar 
mass), such that one O3 dwarf ($\sim$75 $M_{\odot}$) will emit more 
ionizing photons than 25,000 B2 dwarfs ($\sim$9 $M_{\odot}$, Conti et al. 
2008). Therefore, H\,{\sc ii} regions are biased 
towards high mass (O-type) stars with $>$20  $M_{\odot}$
since B stars will produce extremely faint H\,{\sc ii}  regions.

Beyond several Mpc, 
current sensisitivies limit detections of H\,{\sc ii} regions to relatively 
bright examples,  involving several ionizing early O-type stars 
(Pflamm-Altenburg et al. 2007). Still, the H$\alpha$ luminosity of bright
H\,{\sc ii}  regions can be converted 
into the corresponding  number of Lyman continuum  ionizing photons, for 
which the number of equivalent O7 dwarf stars, N(O7V), serves  as a useful reference 
(Vacca \& Conti 1992), as indicated in Table~\ref{table2}. 
Kennicutt et al. (1989) have studied the behaviour of the H\,{\sc ii} 
region luminosity function in nearby spirals and  irregular galaxies. 
Early-type (Sa-Sb) spirals possess a steep luminosity  function, with the 
bulk of massive star formation occuring in small 
regions ionized by one of a few O stars, plus a low cut-off to the 
luminosity function. Late-type spirals and irregulars possess a shallower 
luminosity function, in which most of the massive stars form within large 
H\,{\sc ii} regions/OB complexes, for which 30 Doradus in the LMC serves 
as a useful template. For example, although the LMC contains considerably 
fewer H\,{\sc ii} regions than M31 (SAb), it contains ten H\,{\sc ii} 
regions more luminous than any counterpart in M31 (Kennicutt et al. 1989).

\begin{table}
  \begin{center}
  \caption{Examples of nearby H\,{\sc ii} regions, spanning a range of luminosities 
(adapted from Kennicutt 1984), for an assumed O7V Lyman continuum ionizing flux of 10$^{49}$ ph/s).}
  \label{table2}
  \begin{tabular}{lccccrc}\hline 
  Region & Type & galaxy & Distance & Diameter & L(H$\alpha)$ & N(O7V) \\
         &      &        &  (kpc)    & (pc)     & (erg\,s$^{-1}$ &      \\
  \hline
Orion (M42) &  Classical & Milky Way & 0.5 & 5    &  1$\times 10^{37}$ & $<$1 \\ 
Rosette (NGC 2244)& Classical & Milky Way & 1.5 & 50 & 9$\times 10^{37}$ & 7 \\ 
N66 & Giant & SMC & 60 & 220: & 6$\times 10^{38}$ & 50 \\
Carina (NGC 3372) & Giant    & Milky Way & 2.3 & 300:  &   1.5$\times 10^{39}$ & 120 \\ 
NGC 604 & Giant    & M33 & 800 & 400   & 4.5$\times 10^{39}$ & 320  \\ 
30 Doradus  & (Super)giant& LMC & 50 & 370   & 1.5$\times 10^{40}$ & 1100  \\
NGC 5461 & (Super)giant &M101 & 6400 & 1000: &  7$\times 10^{40}$ & 5000  \\ 
  \hline
  \end{tabular}   
  \end{center}
\end{table}

The integrated nebular H$\alpha$ luminosity of a galaxy is widely used 
as a proxy for the rate of (near-instantaneous) star formation (Kennicutt 1998), 
although conversions into 
total star formation rates (SFR) rely upon the adopted stellar mass 
function and evolutionary models for single and 
binary stars (e.g. Leitherer 2008). In addition, since the 
youngest star forming regions are deeply embedded, the combination of 
gas (H$\alpha$) and dust (24$\mu$m continuum) provide a more complete SFR 
indicator (Calzetti et al. 2007), although the situation is more complicated for 
galaxies with low SFR (e.g. Pflamm-Altenburg et al. 2007). In addition,
H$\alpha$-derived star formation  rates differ from FUV continuum 
diagnostics for dwarf galaxies (Lee et al. 2009b), while FUV indicators 
closely match the local ccSNe rate (Botticella et al. 2012).

\subsection{30 Doradus: Template extragalactic giant H\,{\sc ii} region}

\begin{figure}[tbp]
\begin{center}
 \includegraphics[width=10cm]{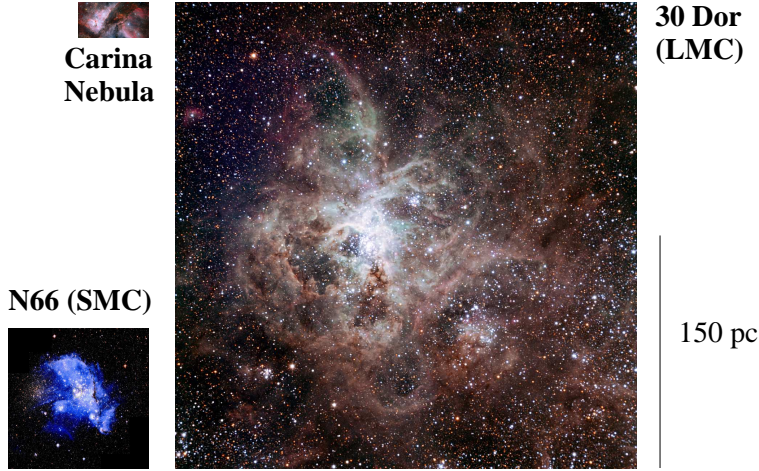} 
 \caption{Three Local Group giant H\,{\sc ii} regions shown on the same physical scale: 
Carina Nebula (Milky Way, ESO/WFI, $60' \times 30'$), 30 Doradus (LMC, ESO/TRAPPIST, 
$20' \times 20'$). N66 (SMC, HST/ACS, $4.7' \times 4.7'$). 
30 Doradus hosts multiple stellar generations (Walborn \& Blades 1997) 
but would only subtend $1.5''$ at a distance of 30 Mpc, so care 
should be taken for characteristic ages of stars within extra-galactic 
H\,{\sc ii} regions.}\label{fig2}
\end{center}
\end{figure}

30 Doradus, the brightest star forming complex within the Local Group, 
provides a useful template for extragalactic `supergiant' H\,{\sc ii} 
regions (Kennicutt et al. 1995, Table~\ref{table2}). The 30 Dor nebula 
is shown in Fig.~\ref{fig2} and spans an angular size of 
$\sim 15' \times 15'$, corresponding to a linear 
scale of 220 $\times$ 220 pc at the distance of the LMC. Consequently, 
individual stars may be studied in detail (e.g. Evans et al. 2011). 
Walborn \& Blades (1997) identified five distinct spatial structures 
within 30 Dor, (i) the central 1--2 Myr cluster R136; (ii) a surrounding 
triggered generation embedded in dense knots ($<$ 1 Myr); (iii) a OB 
supergiants spread throughout the region (4--6 Myr); (iv) an OB 
association to the southeast surrounding R143 ($\sim$5 Myr); (v) an older 
cluster containing red supergiants to the northwest (10--20 Myr). 
30 Dor would only subtend $1.5''$ at a distance of 30 Mpc, so care 
should be taken for nebular-derived ages of stars within extra-galactic 
H\,{\sc ii} regions (e.g. Leloudas et al. 2011).

\section{Environments of supernovae and gamma-ray bursts}

\subsection{H\,{\sc ii} regions and core-collapse SNe}\label{hii}

Turning to studies of the environments of supernovae, locally neither type 
II nor type Ib/c supernovae are associated with ongoing star formation. 
Specifically, Smartt (2009) examined the host environment of a volume 
limited ($cz <$ 2,000 km/s), statistically complete sample of ccSNe, of 
which 0 from 20 type II SN were located in bright H\,{\sc ii} regions. A 
number of type II SN were located in loose associations, with two in older 
clusters (e.g. SN2004am, II-P, in M82), while only 1 of 10 type Ib/c SN 
from Smartt (2009) was in a large star forming region (SN2007gr, Ic, in 
NGC 1058), albeit spatially offset from regions of H\,{\sc ii} emission.

Anderson \& James (2008) took a different approach, studying the 
assocation between ccSNe and H\,{\sc ii} regions within (mostly) bright 
spirals, whose recession velocities extended up to $cz$ = 10,000 km/s. In 
common with Smartt (2009), Anderson \& James (2008) did not find type II 
SNe associated with H\,{\sc ii} regions, concluding that the {\it ``type 
II progenitor population does not trace the underlying star formation''}. 
In contrast, Anderson \& James (2008) found that type Ib, and especially 
Ic ccSNe were spatially coincident with (presumably bright) H\,{\sc ii} 
regions.

\begin{figure}[tbp]
\begin{center}
 \includegraphics[width=6cm]{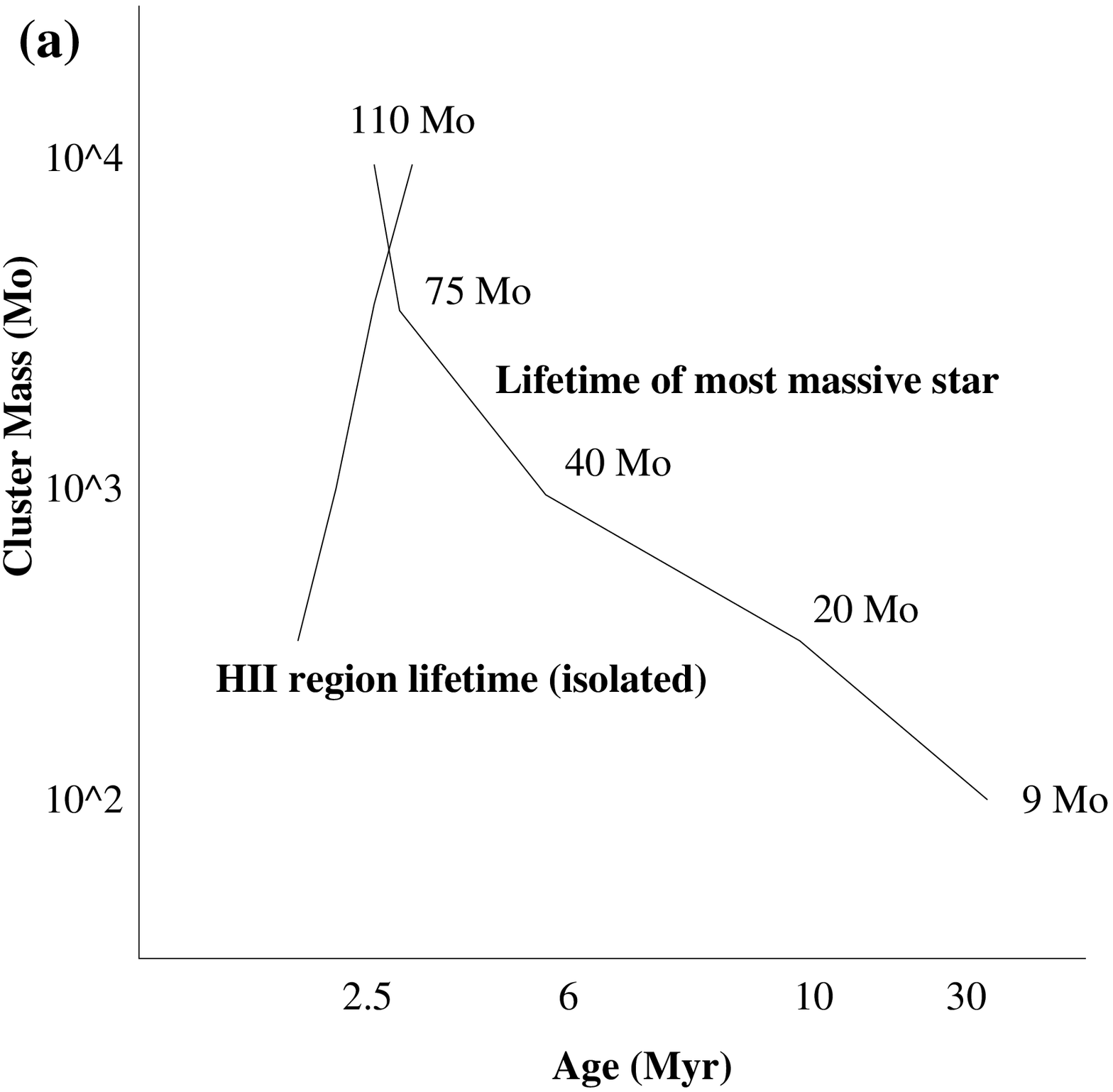} 
 \includegraphics[width=6cm]{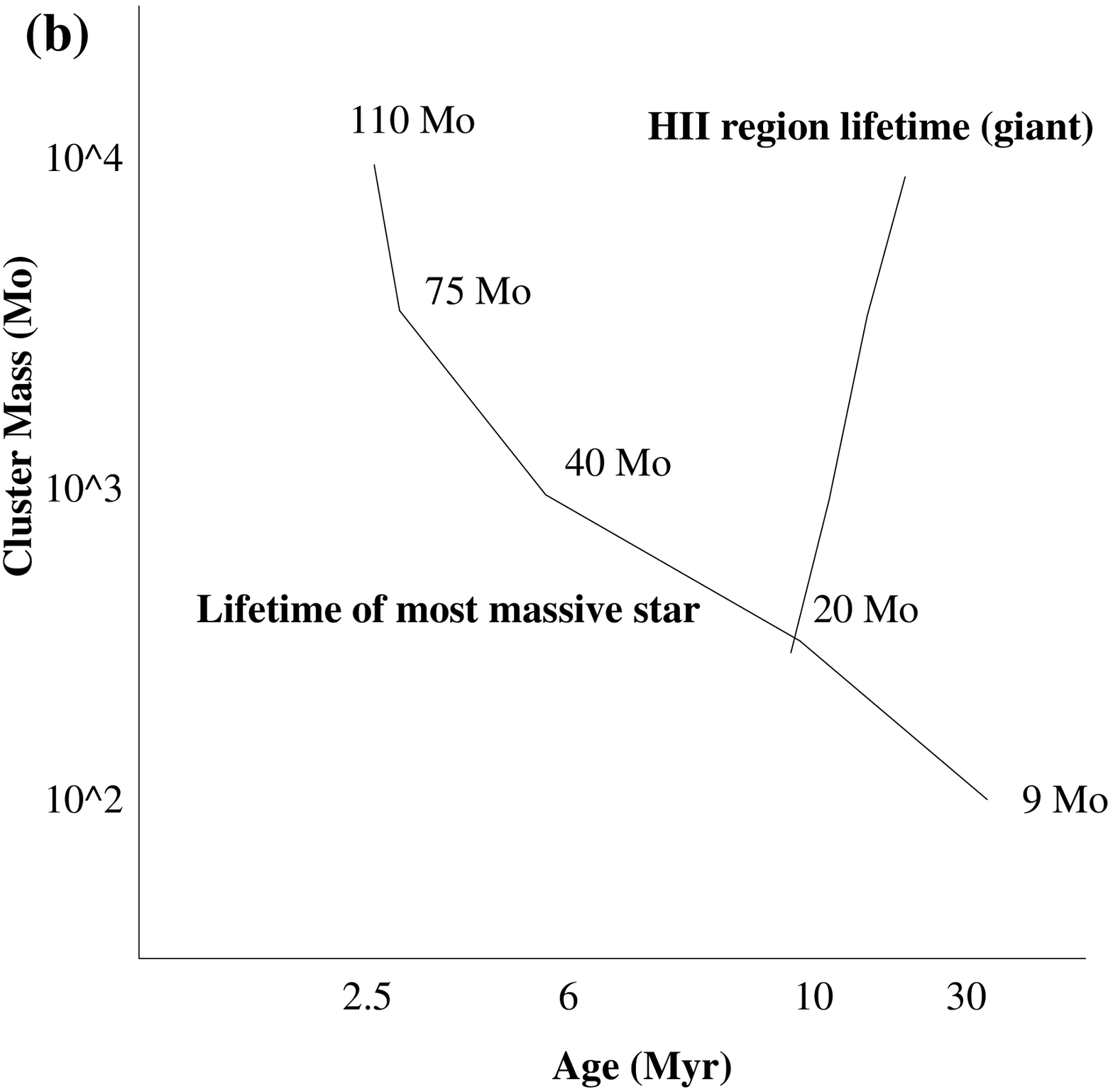} 
 \caption{
(a) Schematic comparing the lifetime of the most massive star in a cluster 
(according to Pflamm-Eltenberg et al. 2007) and isolated H\,{\sc ii} 
regions (adapted from Walborn 2010). Core-collapse SNe should only be 
associated with isolated H\,{\sc ii} regions for very massive progenitors; 
(b) as (a) except for (super)giant H\,{\sc ii} regions, whose 10--20 Myr 
lifetimes should imply an association with ccSNe except for only 
relatively low mass, long lived (type II-P) progenitors.}
\label{fig1}
\end{center}
\end{figure}

Let us consider the typical duration of the H\,{\sc ii} phase in young, 
isolated clusters. Walborn (2010) compared the properties of young star 
clusters, revealing an association with a H\,{\sc ii} region only for the 
first $\sim$2--3 Myr, after which the gas has been dispersed (e.g. 
Westerlund 1, Clark et al. 2005). Therefore, one would {\it not} expect 
ccSNe to be spatially coincident with {\it isolated} H\,{\sc ii} regions 
unless the mass of the progenitor was sufficiently short for its lifetime 
to be comparable to the gas dispersion timescale. This is illustrated in 
Figure~\ref{fig1}(a) where we compare the lifetime of the most massive 
stars in clusters (masses according to Eqn.~10 from Pflamm-Altenberg et 
al. 2007), adopting stellar lifetimes from Ekstr\"{o}m et al. (2012), with 
an estimate of the duration of isolated H\,{\sc ii} regions (adapted from 
Walborn 2010). This naturally explains the lack of any association between 
type II ccSNe and H\,{\sc ii} regions for both Smartt (2009) and Anderson 
\& James (2008).

How, then, can one explain the {\it empirical} association between 
type Ib/c SNe and H\,{\sc  ii} regions in the Anderson \& James (2008) study? 
These either arise from very massive stars, which would be inconsistent with
Smartt (2009), or more likely we have to appreciate that not all massive star 
formation occurs within isolated, compact star clusters.

Late-type spirals and irregulars, which form the majority of Anderson \& James'
host galaxy sample, host large star forming complexes, up to several hundred parsec in 
size, involving (super)giant H\,{\sc ii} regions. These are ionized by successive 
generations of star clusters, separated by a few Myr (Table~\ref{table2}), with a 
total duty cycle of $\geq$10 Myr. Therefore, a massive star exploding within such
an environment as a SN after 5--10+ Myr would still be 
associated  with a bright H\,{\sc ii} region,  
as illustrated in Fig.~\ref{fig1}(b), even if its natal star 
cluster had cleared the gas from its immediate vicinity. Resolving the location of
the ccSNe within the region would be especially difficult at larger distances.
Recall that the average distance of galaxies within the Anderson \& James (2008) sample
was $\sim$32 Mpc, and that their study was based upon moderate resolution
ground-based H$\alpha$ imaging.
Typically higher spatial resolution datasets were employed by Smartt (2009), which together
with a lower host distance ($\sim$27 Mpc maximum for an adopted $H_{0}$ = 75 km/s/Mpc)
enabled a higher spatial inspection of the SN environment (recall Fig.~\ref{fig2}). 

\subsection{Wolf-Rayet stars and ccSNe}

Main-sequence O stars may preceed ccSNe by up to 3--10 Myr, whereas 
Wolf-Rayet stars, their evolved descendents, should preceed (type Ib/c)
SNe by a timescale that is an order of magnitude shorter (Crowther
2007). Therefore, comparisons between the environment of Wolf-Rayet 
stars and ccSNe provide information upon whether the former are plausibly 
the parent population of the latter (e.g. Leloudas et al. 2010), 
since lower mass close binaries might dominate type Ib/c SNe 
statistics (Smith et al. 2011). In 
addition, we can also compare the environment of Wolf-Rayet stars in their 
host galaxies (e.g. associated H\,{\sc ii} regions) with those of ccSNe. 

Since the Wolf-Rayet content of the Milky Way is highly incomplete due to 
foreground interstellar dust, let us instead turn to local external galaxies. 
From the LMC Wolf-Rayet catalogue of Breysacher et 
al. (1999), 58\% (78/134) lie within OB associations, while 83\% (112/134) 
lie within catalogued H\,{\sc ii} regions. Of course, the LMC is not 
particularly representative of the local star forming galaxy population, 
since star formation is largely confined to several giant H\,{\sc ii} 
regions (Kennicutt et al. 1995) and faint H\,{\sc ii} 
regions would not necessarily be identified in distant galaxies. Neugent 
\& Massey (2011) have re-assessed the Wolf-Rayet content of M33 (Scd), 
revealing a total of 206 stars which they argue is complete to $\sim$5\%. 
As for the LMC, the majority of WR stars (80\%) reside in OB associations. 
Further afield, a comparison between the recent 
Wolf-Rayet photometric survey of the Scd spiral NGC 5068 (Bibby 
\& Crowther 2012) with H$\alpha$ images reveals that 50\% of the 
Wolf-Rayet candidates lie in bright or giant H\,{\sc ii} regions, while 
25\% are associated with faint H\,{\sc ii} regions and 25\% lie away from 
any nebulosity. 

Recalling Sect.~\ref{hii}, type Ib/c SN are rarely associated with H\,{\sc 
ii} regions. This suggests that most do {\it not} result 
from young, massive Wolf-Rayet stars, arising instead from lower mass 
close binaries in which the ccSNe arises from the H-deficient, 
mass-losing primary (Fryer et al. 2007). In contrast, the preference of 
type Ic SNe for H\,{\sc ii} regions suggest that massive 
Wolf-Rayet stars {\it are} realistic progenitors in such instances. Close 
binaries in such a scenario ought to mimic the masses, and in turn, the
lifetimes of type II ccSNe progenitors, which positively shy away from H\,{\sc ii} regions.


\begin{table}
  \begin{center}
  \caption{Summary of expected association between H\,{\sc ii} 
regions and ccSNe/long GRBs in different host galaxies (following 
Kennicutt et al. 1989, Gieles 2009).}
  \label{table3}
  \begin{tabular}{lccccrc}\hline 
  Host & SFR & Cluster range & Characteristic      & SN-H\,{\sc ii} & Example\\
       &  & ($M_{\odot}$)  & H\,{\sc ii} region & association?        & \\
  \hline
Spiral (Sab) & Low & 10$^{2-4}$ & Isolated & No (all types)      & M31 \\
Spiral (Scd) & High & 10$^{2-6}$& Giant   & Yes (Ib/c), No (II-P) & M101\\
Irr    & Low & 10$^{2-4}$  & Isolated & No (all types)     & SMC \\
Irr    & High & 10$^{2-6}$& Giant & Yes (Ib/c, GRB), No (II-P) & NGC 1569\\ 
  \hline
  \end{tabular}   
  \end{center}
\end{table}

\subsection{ccSNe, long GRBs and host galaxy types}

Mindful of the spatial resolution issue, let us turn to the Kelly et al. 
(2008) study of SNe locations with respect to the continuum $g'$-band 
light from their low redshift ($z<0.06$) host galaxies. Kelly et al. 
revealed that Ic SNe are much more likely to be found in the 
brightest regions of their hosts than Ib or II SNe. An earlier 
analysis of high redshift galaxies by Fruchter et al. (2006) revealed that 
long GRBs ($<z>$ = 1.25) were also strongly biased towards the brightest pixel 
of their hosts, in contrast to core-collapse SNe ($<z>$ = 0.63, presumably 
mostly type II-P) which merely traced the light from their hosts. Kelly et 
al. (2008) concluded that if the brightest locations correspond to the 
largest star-forming regions, type Ic SNe (and long GRBs) are restricted 
to the most massive stars, while type Ib and especially type II-P SNe are 
drawn from stars with more moderate masses, results in common with 
Anderson \& James (2008).

However, one significant difference between the low-redshift SN H$\alpha$ 
study of Anderson \& James (2008) and the high-redshift GRB study of 
Fruchter et al. (2006) is that hosts of the former are relatively high 
mass, metal-rich spirals, while those of the latter are low mass, 
metal-poor dwarfs. In normal disk galaxies the number of stars forming across the mass distribution
of star clusters is relatively flat, 
albeit with a cut-off linked to the star formation intensity (Gieles 2009).
The star cluster mass function is repeated in nearby dwarf galaxies (Cook
et al. 2012), but galaxy-wide triggers may induce intense, concentrated bursts of 
star formation, leading to disproportionately numerous massive star 
clusters (Billett et al. 2002)\footnote{Of course, not all dwarf galaxies are starbursting. Within the 
local volume ($<$11 Mpc) only a quarter of the star formation from dwarf 
galaxies is formed during starbursts (Lee et al. 2009a)}. We have attempted to set out 
the potential association between H\,{\sc ii}  regions, ccSNe and long GRBs in 
Table~\ref{table3} for star forming
spirals and irregulars, based upon the above arguments, although 
exceptions are anticipated (and subject to uncertainties regarding
the main progenitors of type Ib/c SNe).

Relatively massive, metal-rich galaxies would represent the primary site of all star 
formation for the sample of Fruchter et al. (2006), resulting in (type 
II-P) ccSNe unassociated with the brightest regions in their hosts. Yet, 
when localised starburst activity does occur, it is very intense (Billett
et al. 2002), leading  to very massive clusters, and in turn large numbers of 
high mass, metal-poor stars, a subset of which would be progenitors of the long GRBs 
witnessed by Fruchter et al. (2006).


\section{Upper Mass Limit}

The lower limit to the mass of stars is relatively well known (e.g. 
Burrows et al. 1993), yet establishing whether there is a corresponding 
upper mass limit has proved elusive (Massey 2011). In part, this is 
because obtaining robust masses for VMS is extremely challenging, and 
in part because of the scarcity of star clusters that are sufficiently 
nearby, young and massive for their most massive stars to be studied in 
detail. Up until recently, a mass limit of $\sim 150 M_{\odot}$ has been 
commonly adopted, based upon a near-IR photometric study of the Arches 
cluster (Figer 2005). However, it is well known that the temperature of 
hot, massive stars is rather insensitive to optical/IR photometry. 
Spectroscopic analysis is required for robust temperatures and in turn 
luminosities, from which stellar masses are derived.

\subsection{R136 stars}

The situation is especially difficult for the brightest main-sequence 
members of the most massive young clusters, which possess unusual 
(emission line) spectral morphologies, reminiscent of Wolf-Rayet stars 
(e.g. Drissen et al. 1995). The mass-luminosity relationship for 
main-sequence VMS is relatively flat, $L \propto M^{1.5}$ (e.g. Crowther 
et al. 2012), so inferred masses are particularly sensitive to 
temperature, $M \propto T_{\rm eff}^{8/3}$. Recent advances in atmospheric 
models for stars with dense stellar winds has led to an upward revision to 
the temperatures of such stars, by $\sim$25\%, corresponding to as much as 
an 80\% increase in the resulting mass. Fortunately, several very massive, 
double-lined eclipsing binaries have been identified within the past few 
years, including the Wolf-Rayet binary NGC 3603 A1 (Schnurr et al. 2008), 
permitting an independent check on spectroscopic results for similar 
systems.

R136, the central ionizing cluster of 30 Dor, has both a very high stellar 
mass ($\sim$55,000 $M_{\odot}$) and a sufficiently young age (1--2 Myr) 
for its most massive stars not to have undergone core-collapse. Previous 
estimates of their stellar masses, based on conventional O star 
calibrations, implied 120 -- 155 $M_{\odot}$ (Massey \& Hunter 1998). 
Schnurr et al. (2009) searched for close binaries among the visually 
brightest members, but none revealed radial velocities, with the possible 
exception of R136c. Still, their near-IR integral field datasets provided 
spatially resolved spectroscopy of individual stars within R136, which, 
together with archival UV/optical spectroscopy and AO-assisted photometry 
permitted a reassessment of their stellar masses. Spectroscopic analyses 
together with new evolutionary models for VMS enabled Crowther et al. 
(2010) to revise their (current) stellar masses upward to 135--265 
$M_{\odot}$. Initial masses of 165--320 $M_{\odot}$ were inferred, 
adopting standard main-sequence mass-loss rates for VMS (Vink et al. 2001) 
which closely matched spectroscopically-derived values, and were
reinforced by the close agreement between spectroscopic and
dynamical masses obtained for NGC 3603-A1. Overall, 
R136 supports the trend that higher (initial) mass stars reside within the 
most massive star clusters set out 
by Weidner \& Kroupa (2006). However, statistics of high mass clusters for 
which accurate stellar masses have been determined remain very poor.

\subsection{Pair instability supernovae}

Based upon their re-assessment of the most massive stars in R136 and other 
young, high mass clusters (Arches, NGC 3603), Crowther et al. (2010) 
concluded that their stellar content was consistent with a revised upper 
mass limit of $\sim 300 M_{\odot}$. Regardless of the physical origin of 
this limit, such high initial masses raise the prospect of extremely 
luminous core-collapse SNe (Waldman 2008) or even pair-instability SNe 
(Heger \& Woosley 2002). Models have recently been calculated for the 
post-main sequence evolution of VMS spanning a range of metallicities (N. 
Yusof, these proc.). From these, would appear that the VMS in R136 will 
end their lives as core-collapse SNe, with lower metallicity (SMC-like) 
required to reduce mass-loss rates sufficiently for pair-instability SNe, 
as has been proposed for SN 2007bi (Gal-Yam et al. 2009). However, details 
remain very sensitive to mass-loss prescriptions for the post-main 
sequence evolution (e.g. Crowther et al. 2012).

\acknowledgements{I am grateful to financial support from the Royal 
Society, IAU and local organizers, enabling participation in the 
Symposium. Thanks also to Raphael Hirschi and Lisa Yusof for providing 
results of evolutionary models for VMS prior to publication, plus
Mark Gieles for helpful discussions.}




\end{document}